\documentclass[twocolumn,prl,preprintnumbers,amsmath,amssymb]{revtex4}

\usepackage{xspace}
\usepackage{graphicx}
\usepackage{dcolumn}
\usepackage{bm}

\def\beq{\begin{equation}}
\def\eeq{\end{equation}}

\def\bea{\begin{eqnarray}}
\def\eea{\end{eqnarray}}

\newcommand{\roughly}[1]%
    {{\mathrel{\raise.3ex\hbox{$#1$\kern-.75em\lower1ex\hbox{$\sim$}}}}}

\newcommand{\GeV}{\ensuremath{\mathrm{~GeV}}}

\newcommand{\pb}{\,{\rm pb}}
\newcommand{\Eq}[1]{Eq.~\ref{#1}}

\def\Zp{Z^\prime}
\def\Wp{W^\prime}
\def\Afb{A_{FB}^{t}}

\def\={\,=\,}

\begin{document}

\preprint{MCTP-11-18}
\preprint{CERN-PH-TH/2011-085}

\title{Top quark asymmetry and dijet resonances}

\author{Sunghoon Jung}\email{jungsung@umich.edu}
\author{Aaron Pierce}\email{atpierce@umich.edu}
\affiliation{Michigan Center for Theoretical Physics, Department of Physics,
University of Michigan, Ann Arbor, MI 48109}

\author{James D. Wells}\email{jwells@umich.edu}
\affiliation{Michigan Center for Theoretical Physics, Department of Physics,
University of Michigan, Ann Arbor, MI 48109}
\affiliation{CERN Theoretical Physics (PH-TH), CH-1211 Geneva 23, Switzerland}

\date{\today}

\begin{abstract}
CDF recently reported an anomaly in the $m_{jj}$ distribution of dijet events produced in association with a $W$ boson.
A single $u-t-V$ flavor changing coupling can contribute to the $m_{jj}$ anomaly while being consistent with other resonance searches. Furthermore, it gives a potential explanation of the observed forward-backward asymmetry in top quark production.
\end{abstract}
\maketitle

\section{Introduction}
The CDF collaboration recently released an analysis of the $m_{jj}$ spectrum in  a sample of $\ell {\not}E_{T} jj$ events~\cite{Aaltonen:2011mk}.  The spectrum displays the expected $m_{jj}$ peak at the $W/Z$ mass, but also has a feature near 150 GeV, the significance of which is estimated to be roughly 3$\sigma$.
Additionally, the CDF collaboration recently reported on the asymmetry in top quark production $\Afb$ \cite{Aaltonen:2011kc,cdf:10436}. Focusing on the high-energy region where new physics effects might be expected to be most obvious, CDF measured
$A_{FB}^+ = 0.475 \pm 0.114$	
where $A_{FB}^{+}$ is the asymmetry of top production in the $t \bar{t}$ rest frame restricted to $m_{t \bar{t}} > 450$ GeV.  For comparison, the Standard Model(SM) predicts $A_{FB}^{+}  = 0.088 \pm 0.013$ \cite{Aaltonen:2011kc}. This measurement follows inclusive measurements of the forward-backward asymmetry~\cite{cdf:afb,Aaltonen:2008hc,abazov:2007qb}, which have also consistently yielded large values.

Previous work on the asymmetry posited an explanation for the top quark asymmetry in terms of a new flavor changing boson with mass in the 150--160 GeV range \cite{Jung:2009jz,Jung:2011zv}.  Given the coincidence of mass scales, it is natural to speculate on a common origin for these anomalies (see also \cite{Buckley:2011vc,Nelson:2011us}). Attempts to address the anomaly with a flavor-conserving hadronic $Z^{\prime}$ include refs. \cite{Buckley:2011vc,Yu:2011cw,Wang:2011uq,Anchordoqui:2011ag}.

Here we examine the possibility that these anomalies are indeed related.  In particular, we investigate whether the same particle and same coupling can be responsible for both signals.  As in refs.~\cite{Jung:2009jz,Jung:2011zv},  the $\Afb$ result is explained by a $u-t-V$ coupling, with $V$ a new vector boson.   We demonstrate that this coupling unavoidably contributes to the $m_{jj}$ excess.

\section{New flavor conserving gauge bosons and dijet constraints}

One might think that any vector boson with mass near 150 GeV and appreciable couplings to the SM would already be excluded.  However, as recently reviewed in \cite{Jung:2011zv}, there is room for a light flavor-conserving $\Zp$ that couples exclusively hadronically.  In fact, the strongest published bounds on a $\Zp$ in this mass range are from the UA2 experiment \cite{Alitti:1990kw,Alitti:1993pn}.  The reason is that the gluon parton distribution function rises sharply at low Bjorken-$x$, resulting in an insurmountable QCD dijet background at the Tevatron.

\begin{figure}
\includegraphics[width=0.45\textwidth]{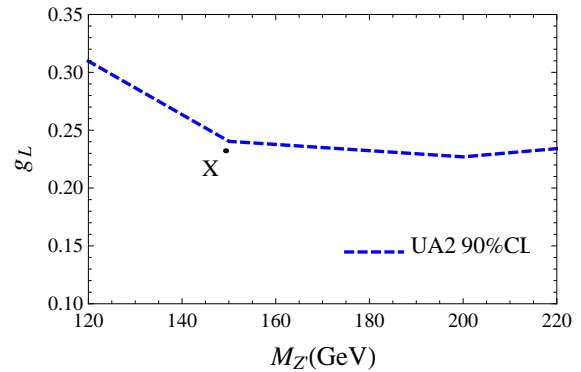}
\caption{Bounds in the $\{ M_{Z'}, g_{L} \}$ plane arising from
dijet resonances at UA2 \cite{Alitti:1990kw,Alitti:1993pn}. Point $X$ is discussed in the text.}
\label{fig:dijet}
\end{figure}

We assume that the $\Zp$ has coupling $g_L$ to left-handed quarks and vanishing coupling to right-handed quarks \footnote{Couplings to left-handed up and down-type quarks should be the same for any new gauge boson that does not have substantial mixing with the Standard Model gauge group.}. This choice should maximize the $W \Zp \to Wjj$ signal while minimizing other signals of the resonance. In Fig.~\ref{fig:dijet}, we show bounds from the UA2 experiment, and mark a point $X$. This point has a mass that could explain the $m_{jj}$ anomaly and has the maximum coupling allowed consistent with UA2 bounds.  Such a model produces an excess of about $160$ events in a mass peak with $4.3$  fb$^{-1}$, to be compared with the excess of $256 \pm 57$ events observed at CDF \cite{Aaltonen:2011mk}.

We also made an attempt to extract a bound on a flavor-conserving $\Zp$ from the $m_{jj}$ spectrum in $\gamma jj$ events (see \cite{Rizzo:1992sh,Barger:1996kr} in an earlier context, and also \cite{Cheung:2011zt}, where it was also noted that the $Zjj$ final state is likely a less sensitive probe). Extant Tevatron data does not appear to be sensitive to the $\gamma + {\rm dijet}$ search discussed above \footnote{A preliminary analysis of CDF data \cite{cdf:10355} would lead one to an opposite conclusion.  However, a discrepancy of a factor of 20 was found in the plots of the CDF note, which when resolved led to a significant weakening of the bounds.}.

\section{A flavor changing gauge boson}
Given the potential constraints from the dijet searches,  it is natural to consider a flavor changing explanation.  Such models would naively be unconstrained by resonance searches (see refs. \cite{Kilic:2011sr,Sato:2011ui,He:2011ss,Eichten:2011sh,Dobrescu} for other approaches). In previous work \cite{Jung:2009jz,Jung:2011zv}, we pointed out the possibility that a gauge boson with flavor-changing $u-t-V$ coupling could explain the $\Afb$ asymmetry via the $t$-channel exchange of $V$. The best point of the original Abelian model of ref.~\cite{Jung:2009jz} predicted $M_V \sim 160 \GeV$ which coincides with the region where the dijet excess is observed. In general, the non-Abelian model of ref.~\cite{Jung:2011zv} can give very similar phenomenology, but does not give rise to a potentially dangerous same-sign top signal, and we use the framework of this second model for our study in this paper.  We comment on the Abelian model as well as alternative possibilities at the end of the section.

The model is described by following fermion interaction Lagrangian \cite{Jung:2011zv}
\begin{eqnarray}
{\cal L} &=& \frac{g_X}{\sqrt{2}} W'^{-}_\mu \Big\{ \, \bar{t}_R \gamma^\mu t_R (-cs) \,+\, \bar{u}_R \gamma^\mu u_R (cs) \nonumber\\
&&  \phantom{\frac{g_X}{\sqrt{2}} W'^{-}_\mu} \,+\, \bar{t}_R \gamma^\mu u_R (c^2) \,+\, \bar{u}_R \gamma^\mu t_R (-s^2) \Big \} \, + \, {\rm h.c.} \nonumber\\
&+& \frac{g_X}{2} \Zp_\mu \Big\{ \, \bar{t}_R \gamma^\mu t_R (c^2-s^2) \,+\, \bar{u}_R \gamma^\mu u_R (s^2-c^2) \nonumber\\
&& \phantom{\frac{g_X}{2} \Zp_\mu \{} \,+\, \bar{t}_R \gamma^\mu u_R (2cs) \,+\, \bar{u}_R \gamma^\mu t_R (2cs) \Big\}
\label{interaction}
\end{eqnarray}
where $c=\cos \theta$ and $s=\sin \theta$. The $\Wp$ is the gauge boson that is responsible for a large $\Afb$ as well as dijet resonance associated with the $W$ boson \footnote{We emphasize that the $\Wp$ notation is a reminder of the underlying $SU(2)_X$ structure -- these states are not electric charged.}. The hadronic $\Zp$ is also present, but its phenomenology is irrelevant for either anomaly discussed here. If the $\theta$ becomes too large, dangerously large same-sign top quark production will re-emerge.

We present two benchmark points of the model in Table \ref{tab:points}. We emphasize this model was presented in an attempt to explain $\Afb$.  Note, the choice of $\cos \theta \neq 1$ combined with the relevant kinematics $M_{\Wp} < M_t$ allow most of $\Wp$s to decay to $u \bar{u}$. The $\Zp$ is light but not constrained by prior experiment \cite{Jung:2011zv}.

\begin{table}
\begin{tabular}{@{\hspace{0.2cm}} c @{\hspace{0.4cm}} c @{\hspace{0.3cm}} c @{\hspace{0.3cm}} c @{\hspace{0.3cm}} c @{\hspace{0.2cm}}}
\hline
\hline
& $M_{W'}$(GeV) & $M_{Z'}$(GeV) & $\alpha_{X}$  & $\cos \theta$\\
\hline
Model A: & 160 & 80 & 0.048 & 0.99 \\
Model B: & 160 & 80 & 0.057 & 0.995 \\
\hline \hline
\end{tabular}
\caption{Benchmark points to be explored below.} \label{tab:points} \end{table}

These two benchmark points are capable of producing a large $\Afb$ while satisfying other bounds on top quark production.  Following the analysis procedure presented in ref.~\cite{Jung:2011zv}, we obtain the results shown in Table \ref{tab:crxTev}, which are consistent with current measurements. Notably, ``faking events" arising from $gu \rightarrow t\Wp $ contribute to the measured top quark $\sigma(t\bar{t})_{\ell j}$ production cross section, making the model fit better with the data. The quoted values of the asymmetry are rest frame parton-level results. In principle these values could be compared with the CDF ``unfolded'' values of $\Afb = 0.158 \pm 0.074$ and  $0.42 \pm 0.16$ quoted in the table.  However, as discussed in \cite{Jung:2011zv} (see also \cite{Gresham:2011pa}), acceptances for these $t$-channel models differ dramatically from the SM, and somewhat smaller values than those quoted in the table would actually be measured.  We can also compare predictions for asymmetries in Model $A$ and $B$ as a function of $\hat{s}$.  We predict $A_{FB}^+$ = 0.42 (A), 0.53 (B), to be compared with $0.475 \pm 0.114$ (CDF), and $A_{FB}^- = 0.08(A), 0.135(B)$, to be compared with $-0.116 \pm 0.153$ (CDF) \cite{Aaltonen:2011kc}.  Here $A_{FB}^+$ ($A_{FB}^-$) corresponds to the observed asymmetry in events with $\sqrt{\hat{s}} > 450$ GeV ($< 450$ GeV). While minor tension exists with the asymmetry measurement at the low $\hat{s}$, on the whole predictions appear consistent with the measured values.

\begin{table} \centering \begin{tabular}{c @{\hspace{0.2cm}} c @{\hspace{0.2cm}} c @{\hspace{0.2cm}} c }
\hline
\hline
 & $\Afb$ & $\sigma(t\bar{t})_{\ell j}$ & $\sigma(t\bar{t})_{\ell \ell}$ \\
\hline
CDF \cite{cdf:afb,Aaltonen:2010ic,cdf:10163,cdf:9890} & $0.158 \pm 0.074$ & $7.22 \pm 0.79 $ pb  & $7.25 \pm 0.92$ pb \\
Model A & 0.25 & 6.9 pb & 5.8 pb \\
Model B & 0.34 & 7.6 pb & 6.4 pb\\
\hline \hline
\end{tabular} \caption{Top production asymmetry, $\Afb$ (rest frame) and apparent top quark production cross sections for points $A,B$ at the Tevatron in the $\sigma(t\bar{t})_{\ell j}$ and
$\sigma(t\bar{t})_{\ell \ell}$ channels.
 Apparent top pair cross sections in the semi-leptonic ($\sigma(t\bar{t})_{\ell j}$) and dileptonic ($\sigma(t\bar{t})_{\ell \ell})$ are obtained by applying CDF selection cuts \cite{Aaltonen:2010ic, cdf:10163, cdf:9890} and by including other faking contributions.  At CDF, a measurement of $\Afb = 0.42 \pm 0.16 $ was also recently made in the dilepton mode~\cite{cdf:10436}.
} \label{tab:crxTev} \end{table}

Finally, we come to the prediction for the excess of the $m_{jj}$ spectrum in $\ell {\not}E_{T} jj$ final states measured at CDF~\cite{Aaltonen:2011mk}.  The dominant contribution in our model comes from $g u \rightarrow t W^{\prime} \rightarrow b W W^{\prime}$ which also contributes to the measured $t\bar{t}$ cross sections.  There are smaller but significant contributions from bottom quark initiated processes (including gluon splitting $g \to b\bar{b}$) $gu \rightarrow b W \Wp$ and $bu \rightarrow W \Wp$. These latter two processes require a mass insertion because the $W$ couples only to left-handed fermions, while the $\Wp$ couples only to right-handed fermions.  Fortunately, the mass insertion occurs on a top quark line, so there is no real suppression due to the large top quark mass.

The extra $b$-jet in the final states is missed some fraction of the time, leading to a signature that contributes to the anomaly.  On the other hand, there is a combinatoric background from incorrectly pairing the $b$-jet with one of the jets from the $W^{\prime}$ decay. This makes the resonance somewhat broader, and slightly non-Gaussian, which is an alternate consistent interpretation of the data. To simulate the $m_{jj}$ spectrum for our model, we use MadGraph~\cite{Alwall:2007st} and smear jet momenta with a Gaussian function of width $\sigma = 0.8 \sqrt{E_T} \, \oplus \, 0.05 E_T$ with $E_{T}$ in GeV (see Table 9.2 in ref.~\cite{Blair:1996kx}). We apply a $K$-factor of $1.03$ for this exclusive two-jet final state -- this number accounts for a NLO correction that partially cancels in the presence of  a jet veto \cite{Baur:1997kz,Campbell:1999ah}. We normalize our $WW/WZ$ sample to CDF expected sample, and apply the same normalization factor to the signal sample. We find that Model A (B) gives an excess of 95 (110) events in the window $110$ GeV $\leq m_{jj}  \leq 180$ GeV, to be compared with an experimentally observed excess of roughly 250 events in the same window.
Although our model points come up short upon first inspection, there are potentially large sources of uncertainty in the comparison of the predicted events with respect to the reported excess due to statistical fluctuations or even due to a few percent error on dijet energy resolution \cite{cmsdoc}.

An Abelian model of $\Zp -u-t$ coupling can also be proposed to account for the excess. As mentioned earlier, the phenomenology of Model point A is essentially identical to the best point considered in~\cite{Jung:2009jz}. A factor of $2$ difference in $\alpha_X$ between the Abelian best point and Model A is simply due to $1/\sqrt{2}$ factor of $\Wp$ interaction in \Eq{interaction}. In contrast, Model B cannot be realized in the Abelian model due to constraints from same-sign dilepton events \cite{Jung:2009jz,AguilarSaavedra:2011zy,cdf:10466}.

Alternatively, we consider a non-Abelian model with $\Wp-d-t$ coupling, which has also been considered in hopes of explaining the top asymmetry \cite{Barger:2010mw,Cheung:2009ch,Cheung:2011qa,Barger:2011ih} (see also \cite{Shelton:2011hq}). By considering both $t\bar{t}$ and dijet measurements, the parameters of $\{ M_{\Wp}, M_{\Zp}, \alpha_X \} =160\GeV, 80\GeV, 0.076$ with a small assumed flavor-diagonal coupling induced by CKM-like mixing can produce $\sigma(t\bar{t})=8.1 \pb, \, \Afb = 0.15$ and about 95 dijet excess events. This constitutes a sizable and tantalizing contribution to the anomaly, but it does not fully explain the excess.

\section{Conclusions}

It is interesting that a model proposed to explain $\Afb$ necessarily gives rise to a resonance near where CDF is declaring an observed excess, and for that reason it is important to investigate fully what the predictions are. If this model approach is correct, a robust prediction is increased single top production from $gu \to t\Wp$ that should and anyway will be pursued by the Tevatron and the LHC experiments. It should be emphasized that our model does not predict the $m_{jj}$ excess as reported, but there are relevant uncertainties in making the comparison between new physics theory and the SM-subtracted measurement.  For our model to be the correct explanation, we must regard the current observation as an upward fluctuation, or alternately, there must be a systematic effect giving rise to part of the observed excess, such as a small offset to the jet energy scale determination.

\section*{ACKNOWLEDGMENTS}

We thank D.~Amidei, R.~Culbertson, T.~Wright, and K.~Zurek for useful discussions and communications. We thank S.~Hewamanage for resolving ambiguities in the CDF note 10355 \cite{cdf:10355}. This work is supported by DOE Grant \#DE-FG02-95ER40899. The work of AP is also supported in part by NSF Career Grant NSF-PHY-0743315.

\bibliography{RapidR}
\bibliographystyle{apsrev}

\end{document}